\documentclass{amsproc}
\pdfoutput=1
\usepackage{amssymb}
\usepackage{graphicx}
\usepackage{bbm}
\newcommand{\one}{\mathbbm{1}}
\newcommand{\id}{\text{id}}
\numberwithin{equation}{section}

\begin{document}

\title{Variable transformation defects}

\author{Nicolas Behr}
\address{Max Planck Institute for Gravitational Physics, Potsdam, Germany}
\curraddr{}
\email{Nicolas.Behr@aei.mpg.de}
\thanks{}

\author{Stefan Fredenhagen}
\address{Max Planck Institute for Gravitational Physics, Potsdam, Germany}
\curraddr{}
\email{Stefan.Fredenhagen@aei.mpg.de}
\thanks{Accepted for publication by the Proceedings of Symposia in
Pure Mathematics of the American Mathematical Society}


\date{November 2011}

\begin{abstract}
We investigate defects between supersymmetric Landau-Ginzburg models
whose superpotentials are related by a variable transformation. It
turns out that there is one natural defect, which can then be used 
to relate boundary conditions and defects in the different models. 
In particular this defect can be used to relate Grassmannian
Kazama-Suzuki models and minimal models, and one can generate rational
boundary conditions in the Kazama-Suzuki models from those in minimal
models. The defects that appear here are closely related to the
defects that are used in Khovanov-Rozansky link homology.
\end{abstract}

\vspace*{-2.5cm}\begin{flushright}
AEI-2012-012
\end{flushright}\vspace*{2.5cm}

\maketitle

\section{Introduction}

Matrix factorisations are a beautiful mathematical subject in the
sense that they are easy to define and still have a lot of interesting
structures. Furthermore they can be used and applied in physics, where
they describe boundary conditions and defects in $N=2$ supersymmetric
Landau-Ginzburg (LG) models (see e.g.\ \cite{Jockers:2007ng} for an overview).

In the simplest setting, a matrix factorisation consists of two quadratic
matrices $p^{0}$ and $p^{1}$ of the same size with polynomial entries whose product is
the identity matrix multiplied by a given potential
$W\in\mathbb{C}[x_{1},\dotsc ,x_{n}]$,
\begin{equation}
p^{0}\cdot p^{1} = W\cdot \one \ , \quad p^{1}\cdot p^{0} = W\cdot
\one\ . 
\end{equation}
An example of that is given by 
\begin{equation}
x^{\ell}\cdot x^{k-\ell} = x^{k} \ ,
\end{equation}
where $p^{0}$ and $p^{1}$ are just polynomials ($1\times 1$
matrices). This example describes B-type boundary conditions in
$N=2$ minimal models~\cite{Kapustin:2002bi,Brunner:2003dc}.

In addition to boundary conditions one can also consider B-type
defects between Landau-Ginzburg models with superpotentials $W$ and
$W'$. They can be described by matrix factorisations of the difference
$W-W'$~\cite{Kapustin:2004df,Khovanov:2004,Brunner:2007qu}. Defects
are very important and useful objects in two-dimensional field theory:
one of their most crucial properties is that they can be fused by
bringing them on top of each other to produce a new defect
\cite{Petkova:2000ip,Brunner:2007qu}. In such a way, defects define an
interesting algebraic structure that turns out to be useful in
analysing symmetries and dualities (see e.g.\ \cite{Frohlich:2006ch}),
and bulk and boundary renormalisation group flows (see e.g.\
\cite{Graham:2003nc,Bachas:2004sy,Brunner:2007ur,Fredenhagen:2009tn})
in such models. As defects can also be fused onto boundaries, they may
be used to relate or to generate boundary conditions. In particular,
if we know defects between different theories, we can generate
boundary conditions in one model from boundary conditions in the other
model by fusion of the defect.

In this work we will analyse defects between LG models with potentials
$W$ and $W'$ that are related by a variable transformation. If these
transformations are non-linear, the two physical theories will be
different. We will see that in such a situation there is one natural
defect that acts in a simple, but non-trivial way on matrix
factorisations. After analysing its properties we will apply it in a
number of examples. In particular we demonstrate how it can be used to
generate matrix factorisations in Kazama-Suzuki models from those in
minimal models.

\section{Variable transformations via defects}

A B-type defect separating two $N=2$ supersymmetric Landau-Ginzburg
models with superpotentials $W$ and $W'$, respectively, can be
described by a matrix factorisation of the difference $W-W'$ of the
potentials \cite{Brunner:2007qu,Carqueville:2009ev}. To be more precise, let $R$ and $R'$ be polynomial
algebras over $\mathbb{C}$, and $W\in R$, $W'\in R'$. A
$(W,W')$-defect matrix factorisation is then a pair $({}_{R}M_{R'},Q)$
where ${}_{R}M_{R'}={}_{R}M_{R'}^{0}\oplus {}_{R}M_{R'}^{1}$ is a
free, $\mathbb{Z}/2\mathbb{Z}$ graded $R$-$R'$-bimodule, and $Q$ is an
odd bimodule map,
\begin{equation}
Q= \begin{pmatrix}
0 & p^{1}\\
p^{0} & 0 
\end{pmatrix} \ ,
\end{equation}
such that $Q^{2} =W\cdot \text{id}_{M} - \text{id}_{M}\cdot W'$. As
$M$ is assumed to be free, $Q$ can be written as a matrix with
polynomial entries.  A B-type boundary condition is a special defect,
for which one side is trivial, e.g.\ $R'=\mathbb{C}$, $W'=0$.

Morphisms between defects $({}_{R}M_{R'},Q)$ and
$({}_{R}\tilde{M}_{R'},\tilde{Q})$ are bimodule maps $\varphi:M \to
\tilde{M}$ with $\tilde{Q}\circ \varphi = \varphi\circ Q$ modulo exact
maps of the form $\tilde{Q}\circ \psi + \psi \circ Q$. Matrix
factorisations are considered to be equivalent if there exist two
morphisms $\phi :M\to \tilde{M}$ and $\psi :\tilde{M}\to M$ such that
$\phi \circ \psi$ and $\psi \circ \phi$ equal the identity map up to
exact terms. Consider e.g.\ $({}_{R}M_{R},Q)$ and $({}_{R}M_{R},S\circ
Q\circ S^{-1})$ for an even isomorphism $S:M\to M$. These
factorisations are then equivalent with the morphisms being $\phi =S$
and $\psi =S^{-1}$. When we write the $\mathbb{Z}/2\mathbb{Z}$
gradation explicitly, the action of $S=\left(\begin{smallmatrix} s^{0}&0\\
0 & s^{1}\end{smallmatrix}\right)$ on $p^{0}$ and $p^{1}$ amounts to
similarity transformations,
\begin{equation}
p^{0} \mapsto s^{1}p^{0} (s^{0})^{-1} \ , \quad p^{1}\mapsto
s^{0}p^{1} (s^{1})^{-1} \ .
\end{equation} 
\smallskip

One of the most interesting properties of defects is that they can be
fused. Physically this means that two defects can be put on top of
each other producing a new
defect~\cite{Petkova:2000ip,Brunner:2007qu}. Mathematically this
amounts to define the tensor product~\cite{Yoshino:1998,Khovanov:2004} of two matrix
factorisations $({}_{R}M_{R'},Q)$ and
$({}_{R'}\tilde{M}_{R''},\tilde{Q})$. As a module this is simply the
graded tensor product
\begin{equation}
M\otimes \tilde{M} = \left(M^{0}\otimes_{R'}\tilde{M}^{0}\oplus
M^{1}\otimes_{R'}\tilde{M}^{1} \right) \oplus  \left(M^{1}\otimes_{R'}\tilde{M}^{0}\oplus
M^{0}\otimes_{R'}\tilde{M}^{1} \right) \ ,
\end{equation}
and the associated module map is 
\begin{equation}
Q\hat{\otimes} \tilde{Q} := \begin{pmatrix}
 0 & \begin{matrix}
p^{1}\otimes \id  & \id \otimes \tilde{p}^{1}\\
-\id \otimes \tilde{p}^{0} & p^{0}\otimes \id
\end{matrix}\\
\begin{matrix}
p^{0}\otimes \id & -\id \otimes \tilde{p}^{1}\\
\id \otimes \tilde{p}^{0} & p^{1} \otimes \id
\end{matrix} & 0
\end{pmatrix}\ .
\end{equation}

For $R'=R$ and $W'=W$, there is a special defect called the identity
defect, which we denote by $({_R}I_{R},{}_{W}\mathcal{I}_{W})$.
Fusing the identity defect onto some defect reproduces the original
defect, it serves therefore as a unit object with respect to the
tensor product. Its precise construction can be found
in~\cite{Khovanov:2004,Kapustin:2004df,Carqueville:2009ev}.
\smallskip

For different superpotentials $W\in R$ and $W'\in S$ there is in general no natural
defect factorisation. On the other hand, if there exists a ring
homomorphism 
\begin{equation}
\phi :R\to S \ , \quad \text{such that}\ \phi (W)=W' \ ,
\end{equation}
then we can naturally map $R$-modules to $S$-modules and vice versa by
extension or restriction of scalars: via the homomorphism $\phi$ the
ring $S$ has a natural $R$-$S$-bimodule structure, ${}_{R}S_{S}$, where
the multiplication from the left is defined via the homomorphism
$\phi$. Given a right $R$-module $M_{R}$ we can then map it to a right
$S$-module by
\begin{equation}
\phi^{*}:M_{R} \mapsto (M_{R})\otimes_{R} ({}_{R}S_{S}) \ ,
\end{equation}    
which describes the extension of scalars from $R$ to $S$.
On the other hand, a left $S$-module ${}_{S}\tilde{M}$ has a natural
$R$-module structure using the homomorphism $\phi$. This restriction
of scalars from $S$ to $R$ can be written as the map
\begin{equation}
\phi_{*}: {}_{S}\tilde{M} \mapsto ({}_{R}S_{S})\otimes_{S} ({}_{S}\tilde{M}) \ .
\end{equation}
$\phi^{*}$ and $\phi_{*}$ act also on module homomorphisms in an
obvious way, so they define functors on the categories of $R$- and $S$-modules.
Notice that $\phi^{*}$ maps free modules to free modules, whereas this
is not guaranteed for $\phi_{*}$. We assume in the following that the
$R$-module ${}_{R}S$ is free, such that $\phi_{*}$ maps free modules to
free modules.

We can apply these functors also to matrix factorisations. In
particular we can apply them to the identity factorisations 
$({}_{R}I_{R},{}_{W}\mathcal{I}_{W})$ and $({}_{S}I_{S},{}_{W'}\mathcal{I}_{W'})$ to
obtain two $(W,W')$-defects with $W'=\phi (W)$,
\begin{equation}
({}_{R}I^{A}_{S},{}_{W}\mathcal{I}_{W'}^{A}) = (\phi^{*} ({}_{R}I_{R}),\phi^{*}
({}_{W}\mathcal{I}_{W})) \ , \quad 
({}_{R}I^{B}_{S},{}_{W}\mathcal{I}_{W'}^{B}) = (\phi_{*} ({}_{S}I_{S}),\phi_{*}
({}_{W'}\mathcal{I}_{W'})) \ . 
\end{equation}
We now claim that these two defects are actually equivalent. To show
this we take the first defect and fuse the identity defect ${}_{S}I_{S}$ from the
right, and compare it to the second defect onto which we fuse the identity
defect ${}_{R}I_{R}$ from the left. As a module we obtain
\begin{equation}
({}_{R}I^{A}_{S}) \otimes_{S} ({}_{S}I_{S}) \cong  ({}_{R}I_{R})\otimes_{R}
({}_{R}S_{S}) \otimes_{S} ({}_{S}I_{S}) \cong  ({}_{R}I_{R}) \otimes_{R} ({}_{R}I^{B}_{S}) \ .
\end{equation}
Since
\begin{equation}
\left({}_{W}\mathcal{I}_{W} \otimes_{R} \id_{S} \right) \hat{\otimes}
\left({}_{W'}\mathcal{I}_{W'} \right) = 
\left({}_{W}\mathcal{I}_{W} \right) \hat{\otimes} \left(\id_{S}
\otimes_{S}\, {}_{W'}\mathcal{I}_{W'} \right) \ ,
\end{equation}
also the factorisations agree, so that we indeed find that
these two defects are equivalent. We call them
$({}_{R}I_{S},{}_{W}\mathcal{I}_{W'})$. 

By a similar consideration as above we see that when we fuse
$({}_{R}I_{S},{}_{W}\mathcal{I}_{W'})$ to the left, it acts by the functor $\phi
^{*}$, whereas it acts by the functor $\phi_{*}$ when we fuse it to
defects to the right. Thus we have a very simple description for the
fusion result for this defect. Analogously we can construct the defect
$({}_{S}I_{R},{}_{W'}\mathcal{I}_{W})$.
\smallskip

Let us explicitly describe how the defect
$({}_{R}I_{S},{}_{W}\mathcal{I}_{W'})$ acts by fusion.  
First consider the (simpler) fusion to the left on a defect
$({}_{R'}M_{R},Q)$. For a rank $2m$ free $R'$-$R$-bimodule
${}_{R'}M_{R}$ we can think of $Q$ as a $2m\times 2m$ matrix with
entries $Q_{ij}$ in $R'\otimes_{\mathbb{C}}R$. Fusing
$({}_{R}I_{S},{}_{W}\mathcal{I}_{W'})$ onto this defect from the right, we obtain a
free $R'$-$S$-module of rank $2m$, and a matrix $\tilde{Q}$ with entries
$\tilde{Q}_{ij}= (\id\otimes \phi) (Q_{ij})$, i.e.\ we just replace
the variables of $R$ by the variables of $S$ via the map $\phi$.

We now assume that ${}_{R}S$ is a finite rank free $R$-module,
\begin{equation}
\rho : R^{\oplus n} \xrightarrow{\sim} {}_{R}S \ .
\end{equation}
With the help of the $R$-module isomorphism $\rho$ we can then explicitly describe
how the defect $({}_{R}I_{S},{}_{W}\mathcal{I}_{W'})$ acts by fusion to the right
on a defect $({}_{S}M_{S'},Q)$. If ${}_{S}M_{S'}$ is free of rank
$2m$, then $Q$ can be represented as a $2m\times 2m$ matrix with entries $Q_{ij}\in
S\otimes_{\mathbb{C}}S'$. After the fusion we have a $R$-$S'$-module of
rank $2mn$, and each entry $Q_{ij}$ is replaced by the $n\times n$
matrix that represents the map $\rho^{-1} \circ Q_{ij}\circ \rho$
(where we tacitly extend $\rho$ to mean $\rho \otimes \id_{S'}$).

A particular situation occurs when all $Q_{ij}$ are of the form $\phi
(\tilde{Q}_{ij})$. As $\rho$ is an $R$-module map, the map $\rho^{-1} \circ
Q_{ij}\circ \rho$ can then be represented by the $n\times n$ matrix
$\tilde{Q}_{ij}\cdot \one_{n\times n}$. The resulting defect is
therefore a direct sum of $n$ identical defects.  As an example,
consider the fusion of $({}_{R}I_{S},{}_{W}\mathcal{I}_{W'})$ on
$({}_{S}I_{R},{}_{W'}\mathcal{I}_{W})$. By the arguments above this
fusion results in a direct sum of $n$ identity defects,
\begin{equation}\label{fundrelation}
({}_{R}I_{S},{}_{W}\mathcal{I}_{W'})\otimes({}_{S}I_{R},{}_{W'}\mathcal{I}_{W})
\cong ({}_{R}I_{R},{}_{W}\mathcal{I}_{W})^{\oplus n} \ .
\end{equation}
\smallskip
In the special case that $\phi$ is a ring isomorphism, $\phi :R\to R$,
and $W'=\phi (W)=W$, the construction above leads to symmetry or
group-like defects $G^{\phi}= ({}_{R}M_{R},(\id \otimes \phi)
({}_{W}\mathcal{I}_{W}))$ (which have been discussed
in~\cite{Frohlich:2006ch,Brunner:2007qu}). The fusion of such defects
is particularly simple,
\begin{equation}
G^{\phi} \otimes G^{\psi} \cong G^{\psi \circ \phi} \ ,
\end{equation}
and $G^{\phi}$ is invertible with inverse $G^{\phi^{-1}}$. These
defects therefore form a group.

\section{Examples and applications}

In this section we want to apply the formalism of the foregoing
section to physically interesting examples.

\subsection*{Minimal models}
Let us first look at the one variable case, $R=\mathbb{C}[y]$, and choose
the potential to be $W (y)=y^{k}$. The corresponding Landau-Ginzburg
model describes a minimal model at level $k-2$. Consider now the ring
homomorphism
\begin{equation}
\phi_{1}: p (y) \mapsto p (x^{d})
\end{equation}
that maps polynomials in $R$ to those in $S=\mathbb{C}[x]$. The
transformed potential is $W' (x) =x^{kd}$. We observe that ${}_{R}S$ is
a free $R$-module of rank $d$,
\begin{equation}
\begin{array}{lrcl}
\rho : & R^{\oplus d} & \to & S\\
       & (p_{1} (y),\dotsc ,p_{d} (y)) &  \mapsto &
       \sum_{j=1}^{d} x^{j-1}p_{j} (x^{d}) \ .
\end{array}
\end{equation}
Let us now look at the corresponding defect between these two minimal
models. We consider the explicit construction
$({}_{R}I^{B}_{S},{}_{W}\mathcal{I}_{W'}^{B})$ via $\phi_{*}$. We start
with the identity defect $({}_{S}I_{S},{}_{W'}\mathcal{I}_{W'})$ that is
given by a rank $2$ matrix
${}_{W'}\mathcal{I}_{W'}=\bigl(\begin{smallmatrix}0 & \imath_{0}\\
\imath_{1} & 0 \end{smallmatrix}\bigr)$ with $\imath_{0}= (x-x')$ and
$\imath_{1}= (W (x)-W (x'))/ (x-x')$. Here we denoted by $x'$ the
variable corresponding to the right $S$-module structure. Under the
map $\phi_{*}$ acting on the left $S$-module structure the entry
$\imath_{0}$ is then replaced by
\begin{equation}
\tilde{\imath}_{0} = \begin{pmatrix}
-x' &    &    & y\\
1  & -x' &    &   \\
   & \ddots & \ddots & \\
  & & 1 & -x'
\end{pmatrix} \xrightarrow[\text{transformation}]{\text{similarity}} \begin{pmatrix}
y-x'^{d} & & & \\
& 1 & & \\
& & \ddots & \\
& & & 1
\end{pmatrix} \ .
\end{equation}
We therefore explicitly see that this defect is equivalent to
$({}_{R}I^{A}_{S},{}_{W}\mathcal{I}_{W'}^{A})$ that we obtain from the
identity defect in $y$-variables by expressing one of the variables in
terms of $x$. This defect is related to the generalised permutation
boundary conditions in two minimal
models~\cite{Caviezel:2005th,Fredenhagen:2006qw} by the folding trick.

We now want to apply this defect to matrix factorisations
$({}_{S}M,Q)$ that describe boundary conditions. The elementary
factorisation $x^{\ell}\cdot x^{kd-\ell}$ will be called
$Q_{\ell}^{(x)}$, and correspondingly $Q^{(y)}_{\ell}$ refers to the
$y$-factorisation $y^{\ell}\cdot y^{k-\ell}$. Fusing the defect
$({}_{R}I_{S},{}_{W}\mathcal{I}_{W'})$ to $(S^{\oplus
2},Q_{rd+\ell}^{(x)})$ results in a superposition $\big( R^{\oplus d},\big(
Q^{(y)}_{r}\big)^{\oplus d-\ell}\oplus \big(
Q^{(y)}_{r+1}\big)^{\oplus \ell}\big)$ (for $0\leq \ell \leq
d-1$). The factorisation $Q^{(y)}_{0}$ is trivial, so we see that the
basic factorisation $Q_{1}^{(x)}$ is just mapped to the basic
factorisation $Q_{1}^{(y)}$.

We can also consider defects in minimal models. Of particular interest
are the group-like defects $G^{n}_{(y)}$ \cite{Brunner:2007qu}
that induce the map $y\mapsto \eta^{n} y$. Here $\eta=\exp \frac{2\pi
i}{k}$ such that the potential $y^{k}$ is invariant. Obviously we have
$G^{n}\cong G^{n+k}$, and the group law is just $G^{m}_{(y)}\otimes
G^{n}_{(y)}=G^{m+n}_{(y)}$. As a $(W,W)$-defect matrix factorisation,
$G^{n}_{(y)}$ corresponds to $(y-\eta^{n}y')\cdot \frac{W (y)-W
(y')}{y-\eta^{n}y'}$.  Similarly, $G^{n}_{(x)}$ denotes the group-like
defect corresponding to the map $x\mapsto \exp \frac{2\pi
in}{kd}x$. Given such a defect $G^{n}_{(x)}$ one can ask what happens to
it when we sandwich it between the defects ${}_{R}I_{S}$ and
${}_{S}I_{R}$. Surprisingly the result can again be expressed in terms
of group like-defects, namely
\begin{equation}
({}_{R}I_{S},{}_{W}\mathcal{I}_{W'})\otimes G^{n}_{(x)} \otimes
({}_{S}I_{R},{}_{W'}\mathcal{I}_{W}) \cong \left(G^{n}_{(y)}
\right)^{\oplus d} \ . 
\end{equation}

\subsection*{\mbox{\boldmath$SU (3)/U (2)$} Kazama-Suzuki model}

As a more interesting example we look at a defect between an $SU (3)/U
(2)$ Kazama-Suzuki model and a product of two minimal models. Consider
the two variable polynomial rings $R=\mathbb{C}[y_{1},y_{2}]$ and
$S=\mathbb{C}[x_{1},x_{2}]$, and the ring homomorphism
\begin{equation}
\phi: p(y_{1},y_{2}) \mapsto p(x_{1}+x_{2},x_{1}x_{2}) \ ,
\end{equation}
which replaces the $y_{i}$ by the elementary symmetric polynomials in
the $x_{j}$. The potential in $x$-variables is that of two minimal models,
\begin{equation}
W' (x_{1},x_{2}) = x_{1}^{k}+x_{2}^{k} \qquad (k\geq 4)\ .
\end{equation}
It is symmetric in $x_{1}$ and $x_{2}$ and thus it can be expressed in
terms of the elementary symmetric polynomials leading to the potential
$W (y_{1},y_{2})$ in the $y$-variables such that $\phi (W)=W'$. This
then describes the $SU (3)/U (2)$ Kazama-Suzuki model (see e.g.\
\cite{Gepner:1988wi,Behr:2010ug}).

The $R$-module ${}_{R}S$ is free of rank $2$ with the explicit
$R$-module isomorphism
\begin{equation}
\begin{array}{lrcl}
\rho : & R\oplus R & \to & {}_{R}S\\
       & (p_{1} (y_{1},y_{2}),p_{2} (y_{1},y_{2})) &  \mapsto &
       p_{1} (x_{1}+x_{2},x_{1}x_{2})+ (x_{1}-x_{2})p_{2}
       (x_{1}+x_{2},x_{1}x_{2}) \, ,
\end{array}
\end{equation}
with inverse
\begin{equation}
\begin{array}{lrcl}
\rho^{-1} : & {}_{R}S & \to &  R\oplus R\\
& p (x_{1},x_{2}) & \mapsto &
\Big( p_{s} (x_{1},x_{2})\big|_{y_{i}} , \frac{p_{a}
       (x_{1},x_{2})}{x_{1}-x_{2}}\big|_{y_{i}}\Big) \ ,
\end{array}
\end{equation}
where $p_{s/a} (x_{1},x_{2}) =\frac{1}{2} (p (x_{1},x_{2})\pm p
(x_{2},x_{1}))$, and $|_{y_{i}}$ means to replace in a symmetric
polynomial in $x_{j}$ the elementary symmetric polynomials by the
$y_{i}$.

The $(W,W')$-defect between the Kazama-Suzuki model ($y$-variables)
and the minimal models ($x$-variables) acts on $y$-factorisations
simply by replacing variables. However, given an $x$-factorisation
with matrix $Q$, a matrix element $Q_{ij}$ is replaced by a $2\times
2$ matrix,
\begin{equation}\label{symmetrisation}
Q_{ij} \mapsto \begin{pmatrix}
( Q_{ij})_{s} & (x_{1}-x_{2}) (Q_{ij})_{a}\\
\frac{(Q_{ij})_{a}}{x_{1}-x_{2}} & (Q_{ij})_{s}
\end{pmatrix} \Bigg|_{y_{1},y_{2}} \ .
\end{equation}
As an example consider the boundary condition based on the
factorisation $(x_{1}-\xi x_{2})\cdot \frac{W'
(x_{1},x_{2})}{x_{1}-\xi x_{2}}$ with $\xi=\exp \frac{\pi i}{k}$
(these are the so-called permutation factorisations~\cite{Ashok:2004zb,Brunner:2005fv}). By the
map~\eqref{symmetrisation} the factor $(x_{1}-\xi x_{2})$ is mapped to 
\begin{equation}
(x_{1}-\xi x_{2}) \mapsto \begin{pmatrix}
\frac{1-\xi}{2}y_{1} & \frac{1+\xi}{2} (y_{1}^{2}-4y_{2})\\[1pt]
\frac{1+\xi}{2} & \frac{1-\xi}{2}y_{1}
\end{pmatrix} \xrightarrow[\text{transf.}]{\text{similarity}} 
\begin{pmatrix}
y_{1}^{2}- 2 (1+\cos \frac{\pi}{k+2})y_{2} & 0\\
0 & 1
\end{pmatrix} \, .
\end{equation}
This means that the linear polynomial factorisation in $x$ is mapped to
a polynomial factorisation in the $y$-variables. The interesting fact
is now that both factorisations describe rational boundary states in
the corresponding conformal field theories~\cite{Behr:2010ug}. 
One can go further and consider the $x$-factorisation
\begin{equation}
\big( (x_{1}-\xi x_{2}) (x_{1}-\xi^{3}x_{2})\big) \cdot \frac{W' (x_{1},x_{2})}{(x_{1}-\xi x_{2}) (x_{1}-\xi^{3}x_{2})} = W'
(x_{1},x_{2}) \ .
\end{equation}
The quadratic factor is mapped to 
\begin{align}
(x_{1}-\xi x_{2})&(x_{1}-\xi^{3}x_{2}) \\
\mapsto & \begin{pmatrix}
\frac{1-\xi^{4}}{2} (y_{1}^{2}-2y_{2})- (\xi+\xi^{3})y_{2} &
\frac{1+\xi^{4}}{2} (y_{1}^{2}-4y_{2})y_{1} \\
\frac{1+\xi^{4}}{2} y_{1} & \frac{1-\xi^{4}}{2} (y_{1}^{2}-2y_{2})-
(\xi+\xi^{3})y_{2} \end{pmatrix}\nonumber\\
\xrightarrow[\text{transf.}]{\text{similarity}} &
\begin{pmatrix}
y_{1}^{2}-2 (1+\cos \frac{\pi}{k})y_{2} & 0\\
y_{1} & y_{1}^{2}-2 (1+\cos \frac{3\pi}{k})y_{2}
\end{pmatrix} \ .
\nonumber
\end{align}
Again this factorisation has been identified with a rational boundary
condition in the Kazama-Suzuki model in~\cite{Behr:2010ug}. This
example shows that the variable transformation defect is indeed very
useful to generate interesting matrix factorisations. In a subsequent
publication we will show that with the help of this variable transformation defect,
one also can generate rational defects in Kazama-Suzuki models which
then allow to generate in principle all factorisations corresponding
to rational boundary conditions in these models.

\begin{figure}[hbtp]
\centering
\includegraphics{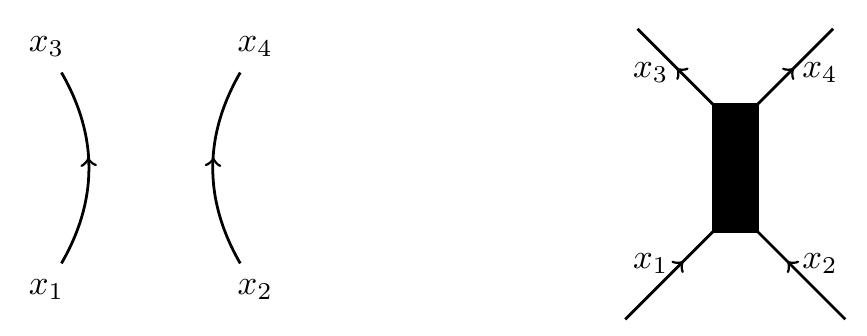}
\caption{Basic building blocks that appear in the resolution of crossings~\cite[figure
9]{Khovanov:2004}: the identity defect ${}_{S'}I_{S}$
in $x$-variables to the left, and the basic wide-edge graph on the right corresponding to $({}_{S'}I_{R})\otimes ({}_{R}I_{S})$ (with ${S=\mathbb{C}[x_1,x_2]}$, ${S'=\mathbb{C}[x_3,x_4]}$).}
\label{fig:KRbblocks}
\end{figure}

\begin{figure}[hbtp]
\centering
\includegraphics{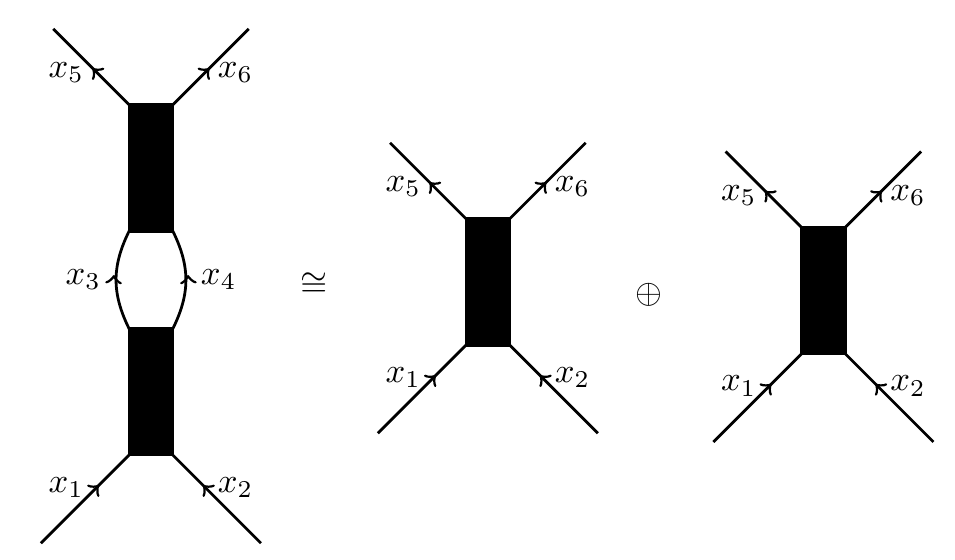}
\caption{One of the fundamental diagram equivalences of \cite[figure 35 and Prop.~30]{Khovanov:2004} (up to grading).}
\label{fig:KRisom}
\end{figure}

The defect considered here actually also appears in the link homology
of Khovanov and Rozansky~\cite{Khovanov:2004}, namely the diagram on
the right in figure~\ref{fig:KRbblocks} corresponds in our language to the defect
$({}_{S'}I_{R})\otimes ({}_{R}I_{S})$ (where $S'=\mathbb{C}[x_3,x_4]$).
The diagram on the left of figure~\ref{fig:KRbblocks} simply
corresponds to the identity defect in $x$-variables. One of the fundamental
equivalences in the link homology displayed in figure~\ref{fig:KRisom} would read
in our notation (with $S''=\mathbb{C}[x_5,x_6]$)
\begin{equation}
\big( ({}_{S''}I_{R})\otimes ({}_{R}I_{S'})\big) \otimes \big( ({}_{S'}I_{R}) \otimes
({}_{R}I_{S})\big) \cong \big( ({}_{S''}I_{R})\otimes
({}_{R}I_{S})\big)^{\oplus 2}\ ,
\end{equation}
which follows immediately from~\eqref{fundrelation}. 
It would be very interesting to also consider the morphisms between
the defects in figure~\ref{fig:KRbblocks} that are needed to formulate the complex of defects
assigned to crossings (see~\cite[figure 46]{Khovanov:2004}) in our
framework, but we leave this for future work.

\subsection*{\mbox{\boldmath$SU (n+1)/U (n)$} Kazama-Suzuki models}

The last example has a beautiful generalisation to a defect between a
$SU (n+1)/U (n)$ Kazama-Suzuki model and $n$ copies of minimal
models. We consider the polynomial rings $R=\mathbb{C}[y_{1},\dotsc
,y_{n}]$ and $S=\mathbb{C}[x_{1},\dotsc ,x_{n}]$, and the potential 
\begin{equation}
W' (x_{1},\dotsc ,x_{n}) = x_{1}^{k}+ \dotsb +x_{n}^{k} \ .
\end{equation}
The ring homomorphism is defined by
\begin{equation}
\phi (y_{j}) = \sum_{i_{1}<\dotsb <i_{j}} x_{i_{1}}\cdot \dotsb \cdot
x_{i_{j}} \ ,
\end{equation}
and it maps the $y_{j}$ to the elementary symmetric polynomials in the
$x_{i}$. It is an old result in invariant theory~\cite[section
II.G]{Artin} that ${}_{R}S$ is a free $R$-module of rank $n!$. To get
an explicit $R$-module isomorphism between ${}_{R}S$ and $R^{\oplus
n!}$, one needs to choose a good basis in $S$. The simplest
choice~\cite{Artin} is to take the $n!$ polynomials given by
\begin{equation}
x_{1}^{\nu_{1}}x_{2}^{\nu_{2}}\cdot \dotsb \cdot x_{n}^{\nu_{n}} \ , \
\text{where} \ \nu_{i}\leq i-1 \ .
\end{equation}
Another possibility with some computational advantages is provided by
the Schubert polynomials $X_{\sigma} (x_{1},\dotsc ,x_{n})$, for which there is one for
each permutation $\sigma$ in the symmetric group $S_{n}$. It was shown
in~\cite{Macdonald} that any polynomial $p (x_{1},\dotsc ,x_{n})$ has a unique
expansion
\begin{equation}
p (x_{1},\dotsc ,x_{n}) = \sum_{\sigma\in S_{n}} p_{\sigma} (x_{1},\dotsc
,x_{n})X_{\sigma} (x_{1},\dotsc ,x_{n}) \ ,
\end{equation}
where the $p_{\sigma}$ are totally symmetric. The map $\rho$ is then
given by
\begin{equation}
\begin{array}{lrcl}
\rho : & R^{\oplus n!} & \to &  S \\
       & \!\!\!\!(p_{\sigma} (y_{1},\dotsc ,y_{n}))_{\sigma \in S_{n}}\! & \mapsto &
\!\!\!{\displaystyle \sum_{\sigma \in S_{n}}} p_{\sigma} (x_{1}+\dotsb +x_{n},\dotsc
,x_{1}\dotsb x_{n}) X_{\sigma} (x_{1},\dotsc ,x_{n})\, .      
\end{array}
\end{equation}

\section{Conclusion}

We have seen that there is a natural defect between Landau-Ginzburg
theories whose superpotentials are related by a variable
transformation. The fusion of this defect onto other factorisations
has an explicit and simple description via the functors $\phi^{*}$ and
$\phi_{*}$ corresponding to extension and restriction of scalars.

The examples have shown that these defects can be used to relate
boundary conditions or defects in different LG models. In particular,
one can use such defects between minimal models and Grassmannian
Kazama-Suzuki models to put into use the knowledge that is already
available for minimal models to obtain factorisations for the
Kazama-Suzuki models. In a such a way one can for example generate all
factorisations corresponding to rational boundary conditions in the
$SU (3)/U (2)$ model, as we will show in a subsequent publication.

In the $SU (3)/U (2)$ example it turns out that the defects discussed
here are crucial to construct factorisations for rational topological
defects. These finitely many elementary defects and their
superpositions form a closed semi-ring that is isomorphic to the
fusion semi-ring. Realising such a finite-dimensional semi-ring (in
the sense that as a semi-group it is isomorphic to a direct product of
finitely many copies of $\mathbb{N}_{0}$) in terms of defect matrix
factorisations reflects the rational structure of the conformal field
theory that is otherwise hard to see in the LG formulation. It would
be interesting to investigate whether the existence of such an
algebraic structure automatically signals an enhanced symmetry in the CFT. 

Finally, we have seen that our defects generate the building blocks of
Khovanov-Rozansky homology, except for the morphisms between defect
building blocks. In other words, one could say that our formulation
provides a physical setup of the Khovanov-Rozansky factorisations as a
sequence of Kazama-Suzuki models separated by defects.  By
generalisation to $SU (n+1)/U (n)$ models, this is also true for the
higher graphs appearing in the MOY calculus~\cite{Murakami}. Our
analysis can therefore be seen as a physical supplement to the recent
results in~\cite{Becker,Carqueville:2011sj}.

\section*{Acknowledgements}
We thank Nils Carqueville, Dan Murfet and Ingo Runkel for useful discussions.

\end{document}